# EasySched : une architecture multi-agent pour l'ordonnancement prédictif et réactif de systèmes de production de biens en fonction de l'énergie renouvelable disponible dans un contexte industrie 4.0

Maroua Nouiri, Damien Trentesaux, Abdelghani Bekrar
LAMIH UMR CNRS 8201, Université Polytechnique Hauts-de-France

Le Mont Houy, 59313, Valenciennes, France.
Corresponding author: maroua.nouiri@uphf.fr

**Résumé :**

L'industrie 4.0 s'accompagne de la prise en compte de contraintes de développement durable. Dans ce contexte, nous proposons une architecture multi-agent pour l'ordonnancement prédictif et réactif coordonné entre des systèmes de production de biens et des systèmes de production d'énergie renouvelable, appelée EasySched. La validation de cette architecture est originale, elle est menée de manière complètement et physiquement distribuée en utilisant des systèmes embarqués en réseau. Cette validation est menée sur une série d'instances inspirées de la littérature. Les résultats montrent que les mécanismes proposés permettent d'adapter la production selon l'énergie renouvelable disponible.

**Mot clés :** production, ordonnancement prédictif, ordonnancement réactif, énergies renouvelables, modélisation multi-agent, Optimisation par essaims particulaires, systèmes embarquées

## 1. Introduction

Les enjeux environnementaux, tels que la réduction de la pollution ou des émissions de dioxyde de carbone ($CO_2$) et la maîtrise des changements climatiques prennent une importance croissante. L'industrie 4.0 se doit d'assurer un équilibre entre un accroissement des besoins énergétiques, et la réduction de l'impact environnemental des ressources énergétiques utilisées (**Cardin et al., 2017), (Trentesaux et al., 2017)**. Le système énergétique mondial subit des transformations en profondeur de par l'avènement des énergies renouvelables. Néanmoins, la disponibilité de ce type d'énergie est fortement variable et difficilement prévisible (**Prabhu et al., 2015**).

Afin de pallier ce problème, il est primordial de mettre en place et d'optimiser une coopération entre la source (les fournisseurs d'énergies) et le client (les usines de production) (**Trentesaux et al., 2016)**. Cette coopération doit être réactive pour pouvoir faire face à des événements difficilement prévisibles, de fréquence et d'importance variées.

Dans ce contexte, nous proposons EasySched, une architecture multi agent pour l'ordonnancement prédictif et réactif de systèmes de production de biens (consommateur d'énergie) en fonction des contraintes issues de systèmes de production d'énergie renouvelables (fournisseurs) dans un contexte Industrie 4.0. L'ordonnancement est en effet une fonction de la production qui doit désormais être abordée en tenant compte d'aspects relatifs au développement durable et à la gestion de l'énergie **(Giret et al., 2015)**. Pour ce faire, un mécanisme de coopération entre ces systèmes de production est proposé. Il concerne le calcul (en prédictif) et l'ajustement (en réactif) de l'ordonnancement de la production de biens vis-à-vis de la consommation d'énergie. La production d'énergie de type renouvelable est difficilement prédictible car fortement dépendante de conditions non contrôlables (par exemple, les conditions météorologiques). L'originalité de cette architecture réside dans le fait que l'élaboration de l'ordonnancement se fait en tenant compte des besoins en production et du niveau dynamique d'énergie renouvelable disponible. La validation de cette architecture a été faite de manière complètement distribuée, en intégrant à la fois du matériel et du logiciel. Dans cet article, et pour illustrer nos propos, deux sources d'énergie (éolienne et photovoltaïque) sont considérées et les systèmes de production considérés sont de type job-shop, la variable d'ajustement sur la consommation étant la vitesse des machines de production. Notre approche ambitionne ainsi d'adapter en temps réel la production, et donc la consommation d'énergie dynamiquement à l'apport énergétique.

Dans ce qui suit, nous présentons en premier lieu un état de l'art relatif à la production sous contrainte énergétique ou en environnement perturbé. Les détails de notre architecture seront présentés dans la section 3. La section 4 présentera les divers tests réalisés et les résultats d'expérimentation. Une conclusion et des perspectives seront fournies dans la section 5.

## 2. État d'art

Nos travaux relèvent de la problématique scientifique suivante : une production de biens doit être ordonnancée de manière prédictive et réactive en tenant compte de l'évolution prévisionnelle et en temps réel de l'énergie renouvelable dont la disponibilité est incertaine. Il existe de nombreuses approches relevant de ce contexte. Sans vouloir être exhaustif, nous présentons dans cette partie un ensemble de références qu'il nous semble important d'étudier et que nous avons regroupées en trois catégories.

Une première catégorie traite de cette problématique de manière complètement statique. Par exemple, **Mikhaylidi et al., (2015)** ont considéré un problème de contrôle des opérations de fabrication avec des prix de l'électricité connus variant dans le temps sur un horizon de planification fini. L'objet est de déterminer la date à laquelle chaque opération doit être effectuée dans l'horizon temporel donné afin de minimiser la consommation totale d'électricité et les coûts liés aux pénalités de retard des opérations. Les auteurs ont proposé une solution de type programmation dynamique. **Gahm et al., (2016)** ont défini un ordonnancement économe en énergie, avec un objectif d'amélioration de l'efficacité énergétique. Pour évaluer leur travail, ils ont pris en compte non seulement les processus et les systèmes de production, mais également les processus et les systèmes d'approvisionnement en énergie. **Fang et al., (2011)** ont présenté un modèle de programmation mathématique du problème d'ordonnancement d'ateliers qui prend en compte la charge de pointe, la consommation d'énergie et l'empreinte carbone associée, en plus du temps de cycle. Le modèle est illustré à l'aide d'une étude de cas simple : un atelier de fabrication où deux machines sont utilisées pour produire une variété de pièces. Le problème d'ordonnancement proposé considère la vitesse de fonctionnement comme une variable indépendante, qui peut être modifiée pour affecter la charge de pointe et la consommation d'énergie. **Gonzalez et al., (2017)** ont suggéré une méta-heuristique hybride associant un algorithme génétique à une nouvelle méthode de recherche locale et une approche de programmation linéaire pour améliorer le coût énergétique d'un ordonnancement d'atelier de type job shop. **He et al., (2015)** ont conçu une méthode d'optimisation économe en énergie en tenant compte de la sélection de la machine et de la séquence de fonctionnement. **Tonelli et al., (2016)** ont proposé un modèle centralisé et distribué pour trouver l'ordonnancement prédictif hors ligne. **Plitsos et al., (2017)** ont défini un système d'aide à la décision (DSS) composé d'un algorithme de recherche local qui offre une optimisation hiérarchique (basée sur plusieurs objectifs) de la planification de la production. L'énergie est considérée sous la forme de contraintes. **Lamy., (2017)** a élaboré un modèle mathématique et une méta-heuristique Greedy Randomized Adaptive Search Procedure (GRASP) pour résoudre le problème de job-shop avec contrainte énergétique. Dans cette catégorie, tous les travaux présentés ont porté sur des modèles d'ordonnancement statique avec optimisation de la consommation d'énergie.

Plusieurs sortes d'aléas peuvent arriver en cours de la production et peuvent être de différentes origines (**Chaari et al., 2014**). Ceci complexifie les problèmes d'ordonnancement. La solution produite doit en effet prendre en compte les perturbations de l'environnement, tout en assurant de bonnes performances. Dans ce contexte, une seconde catégorie de méthodes a été développée pour prendre en compte l'incertitude lors de la résolution du problème d'ordonnancement. **Chaari et al., (2014)** ont suggéré une classification qui positionne les différentes méthodes d'ordonnancement sous incertitudes. Celle-ci comporte trois types : les approches proactives, réactives et hybrides. Cette dernière regroupe les approches prédictives-réactives et les approches proactives-réactives. Les approches prédictives réactives sont constituées de deux phases : la première phase consiste à construire un ordonnancement déterministe hors-ligne ne prenant pas en considération les événements imprévisibles. Par exemple, les opérations prévues sont toutes connues dès le départ, les temps de traitement ont été préalablement déterminés et les machines sont disponibles tout le temps durant l'ordonnancement. Durant la deuxième phase, cet ordonnancement est utilisé en ligne. Il est adapté en temps réel pour tenir compte des perturbations (**Chaari et al., 2014**). Au delà du mécanisme de réactivité qui doit être fourni par la méthode d'ordonnancement, un objectif de durabilité doit aussi être visé dès la phase de conception. Par exemple, **Salido et al., (2016)** se sont concentrés sur le problème de ré-ordonnancement d'ateliers dynamiques où les machines peuvent fonctionner à des vitesses différentes. Les auteurs ont proposé une nouvelle technique basée sur un algorithme génétique pour trouver un ordonnancement avec un temps assurant la minimisation de la consommation d'énergie. **Zhang et al., (2013)** ont défini un nouveau modèle mathématique pour résoudre le problème de job-shop flexible qui prend en compte simultanément la consommation d'énergie et l'efficacité de la planification. Ils ont également proposé une méthode de ré-ordonnancement basée sur un algorithme génétique.

Les travaux précédemment cités proposent des méthodes de ré-ordonnancement en tenant compte de la consommation d'énergie face à une classe de perturbations de type « panne machine » ou « attribution d'une tâche aléatoire ». Cependant, ils ne tiennent pas compte de la variation dynamique de la disponibilité énergétique. La troisième et dernière catégorie de travaux relève de cet aspect. Par exemple **Pach et al., (2015)** ont proposé une méthode qui considère la variation de la consommation d'énergie comme contrainte mais aussi comme étant une source d'incertitude. Les auteurs ont proposé une approche basée sur les champs de potentiel pour activer / désactiver les ressources de manière réactive en fonction de la situation du système de fabrication flexible (FMS) afin de réduire la perte d'énergie. Le contrôle de la consommation électrique globale a alors été introduit durant la fabrication afin de respecter un seuil d'énergie disponible déterminé dynamiquement et a priori inconnu. Dans (**Trentesaux et al., 2017**), une architecture multi-agent prédictive et réactive intégrant consommateurs d'énergie et producteurs d'énergie a été spécifiée, cependant aucune modélisation n'a été proposée. En se basant sur ce travail, **Nouiri et al., (2018)** ont élaboré un modèle MA-EAPSRS (Multi Agent-Energy Aware Production Scheduling and Rescheduling System) en décrivant les différents agents et leurs comportements. Des protocoles de négociation ont été proposés et validés pour calculer de manière prédictive une production durable. Cependant, aucun mécanisme de décision réactive n'a été proposé.

De notre point de vue, quelle que soit la catégorie de contribution concernée, les travaux restent à un niveau de validation très conceptuel, au mieux, en simulation et considèrent très peu une variabilité dans la disponibilité de l'énergie alors que les énergies renouvelables se développent de plus en plus. Il nous semble ainsi important de proposer des algorithmes pour ordonnancer de manière prédictive et réactive des centres de production de bien, consommateurs d'énergie, en tenant compte de contraintes issues de centres de production d'énergie renouvelables. Une validation sur des expérimentations plus réalistes de ces algorithmes est également souhaitable. Ces deux points font l'objet de cet article.

Plus précisément, notre contribution porte sur l'enrichissement, au travers d'une architecture multi-agent expérimentale qu'il est possible de valider de manière complètement distribué, intégrant matériel et logiciel, de l'architecture précédemment proposée MA-EAPSRS. Cet enrichissement porte sur l'intégration de mécanismes réactifs permettant d'adapter la production à l'énergie disponible. Afin de valider cette architecture multi-agent, des expérimentations seront proposées pour planifier en temps réel, la consommation énergétique des usines connectées à un réseau intelligent tenant compte d'une fluctuation de production énergétique. La validation aura ainsi lieu sur les parties prédictives et réactives.

Nous présentons dans la suite tout d'abord le modèle multi-agent support de l'architecture et la contribution scientifique concernant la dimension réactive avant de présenter sa réalisation logicielle et matérielle qui nous permettra de tester et valider les mécanismes de coopération proposés.

## 3. Proposition d'un modèle multi-agent support de l'architecture

### 3.1 Hypothèses de travail

Les décisions prises par les systèmes de production que nous étudions sont des décisions d'ordonnancement prédictif et réactif. Il s'agit de programmer l'exécution d'un ensemble de tâches à des dates spécifiques, tenant compte d'une contrainte énergétique communiquée en temps réel. Les contraintes considérées lors de l'établissement de l'ordonnancement sont présentées **figure 1**. Ces contraintes peuvent être classées selon leur nature déterministe (contrainte de précédence par exemple) ou dynamique, prenant en considération la disponibilité des machines au fil du temps, par exemple.

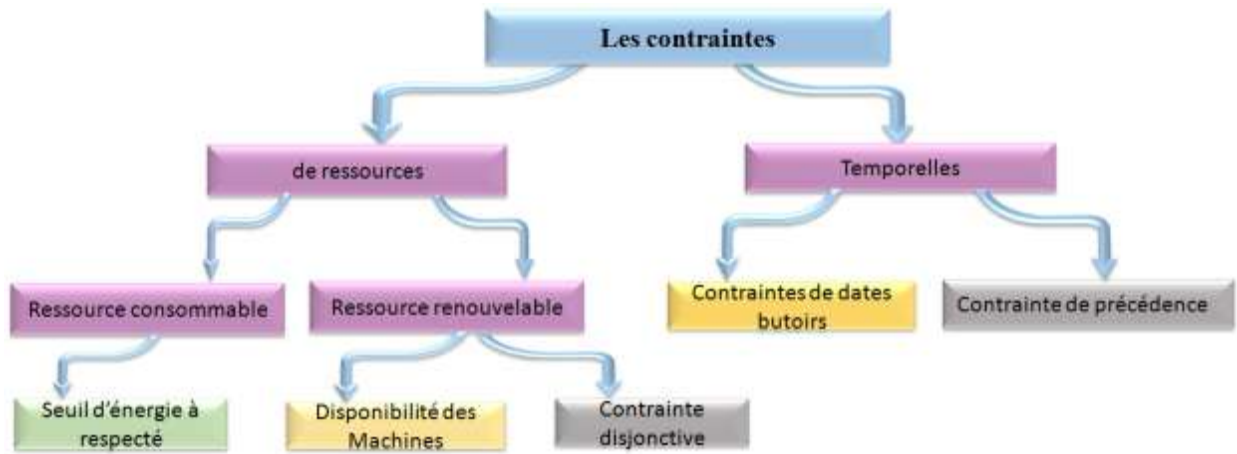

**Figure 1. Contraintes dans un problème d'ordonnancement.**

Notre objectif est d'optimiser la date de réalisation de la production (makespan) tout en assurant l'efficience de celle-ci par la minimisation de la consommation énergétique des ressources de fabrication (soit usine à l'échelle macroscopique, soit machines à l'échelle microscopique intra-usine). Nous ne considérons pas les autres consommateurs d'énergie (systèmes de transport, chauffage, etc.).

La méthode proposée est hybride et comporte deux phases. Elle est basée sur une coopération entre les systèmes consommateurs d'énergie et les systèmes producteurs d'énergie. La phase prédictive de la méthode proposée consiste à calculer hors ligne un ordonnancement pour une consommation d'énergie déterminée (cette phase a été décrite dans (**Nouiri et al., 2018**)). La deuxième phase, réactive, est détaillée dans cet article. Elle a pour objectif l'ajustement, en temps réel, des plans de planification en tenant compte des évènements non prévus. Ces évènements peuvent êtres internes comme les pannes de machines ou externes (ordres urgents).

La **figure 2** illustre les différents types de fournisseurs et consommateurs d'énergie qu'il est possible de considérer, que ce soit au niveau des systèmes de production consommateurs d'énergie (en production manufacturière : single machine, flow shop, ateliers flexibles, etc.) ou au niveau des fournisseurs d'énergie renouvelable (énergie éolienne, photovoltaïque, etc.).

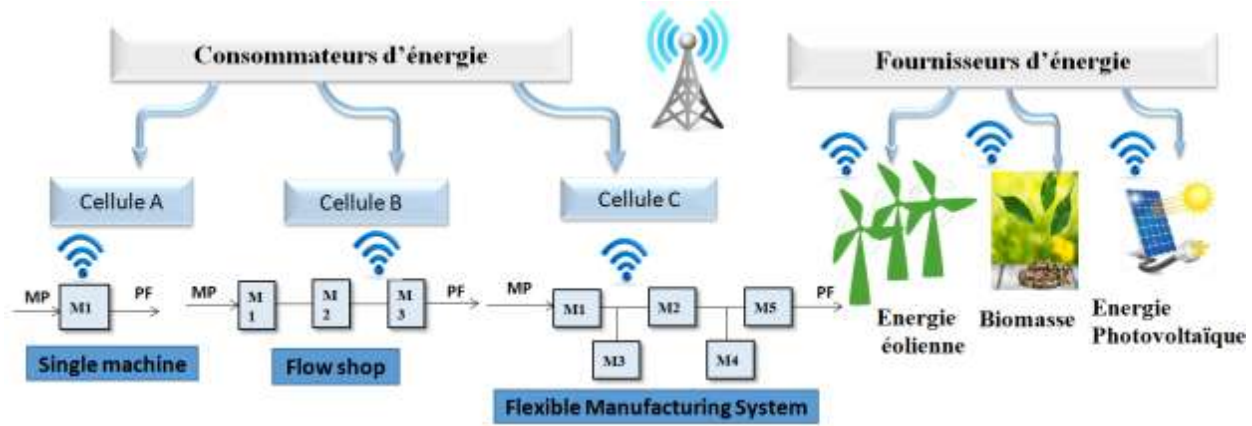

**Figure 2. Différents types de fournisseurs et consommateurs d'énergie.**

Dans le cadre de cet article, nous considérons que les usines sont composées de systèmes de production de type job-shop avec pour chacun, la possibilité de faire varier les vitesses de fonctionnement des machines (ce qui modifie leur consommation énergétique) et deux sources d'énergie uniquement de type renouvelable, éolien et photovoltaïque. Le changement de vitesse des machines fait référence à la variation de leur cadence et non à un arrêt total pendant une période de temps. Dans ce qui suit, nous détaillons les types d'agents, leurs comportements et décrivons les types de perturbations considérées.

### 3.2 Types d'agents et comportements

L'approche multi-agent permet de mettre en œuvre des mécanismes de type prédictif (en intégrant par exemples des algorithmes d'optimisation) et réactif (par interaction dynamique) (**Whitrook et al., 2018**). C'est cette capacité qui a guidé notre choix d'une approche multi-agent. Il existe de nombreuses manières de définir, structurer et articuler les interactions entre agents dans un contexte de couplage entre une phase prédictive et une phase réactive (**Jimenez et al., 2017**). Dans le cadre de nos travaux, l'architecture multi-agent proposée est complètement distribuée afin de faciliter la réactivité face à une variété d'événements imprévus. Elle a été conçue selon la méthodologie de conception Go-Green ANEMONA (**Giret et al., 2017**). Elle est constituée d'agents spécialisés qui représentent d'une part les systèmes de production de bien et d'autre part les fournisseurs d'énergie. Les rôles de ces agents sont articulés selon ces deux phases (prédictives et réactives). Nous associons un agent à chaque usine de production : chaque usine est ainsi représentée par un agent nommé "Agent Ordonnanceur d'usine" (AOU). Chaque fournisseur d'énergie est représenté de manière symétrique par un agent nommé "Agent Ordonnanceur d'énergie" (AOE). Un agent nommé " Agent Contrôleur Energie" (ACE) est également proposé. Son rôle est de contrôler en temps réel les consommations d'énergie des usines connectées. **La figure 3** schématise les rôles des différents agents de l'architecture proposée appelée EasySched (Energy-Aware production SYstem SCHEDuling). Nous détaillerons dans ce qui suit les agents AOU et AOE.

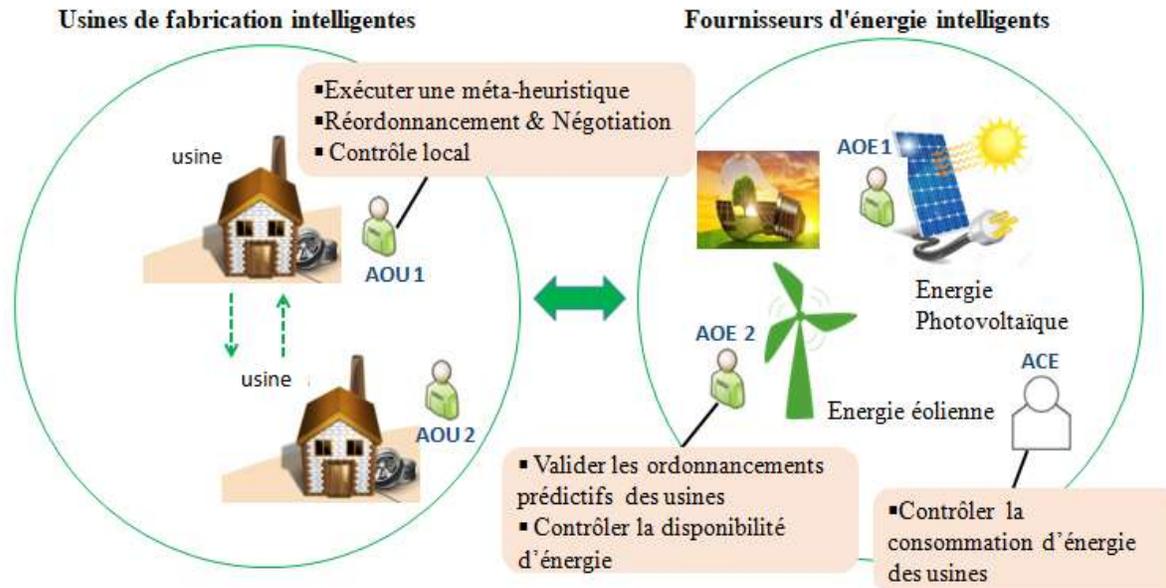

**Figure 3. Rôles des agents dans l'architecture EasySched**

### 3.2.1. Agent Ordonnanceur de fournisseur d'énergie (AOE)

L'Agent Ordonnanceur de fournisseur d'énergie a pour rôle de :

- Valider les ordonnancements prédictifs et les demandes en énergie envoyées par les usines connectées. (hors ligne)
- Contrôler la disponibilité de l'énergie renouvelable. (en ligne)

Le rôle en mode hors ligne consiste à valider les ordonnancements prédictifs. La validation porte sur les demandes en consommation en énergie et leurs capacités à être satisfaites ou non (elles sont dès lors refusées).

Le rôle du mode en ligne est relatif au contrôle et au suivi, en temps réel de la disponibilité de l'énergie renouvelable. Il s'agit d'un comportement cyclique qui permet de collecter les données des capteurs spécifiques à chaque type de ressource (par exemple, de température, d'humidité, de luminosité, du vent, etc.). Les données ainsi collectées seront utilisées pour alimenter les processus de prise de décision de ré-ordonnancement. Des boucles de rétroaction plus complexes peuvent être créées sur la base des données sensorielles obtenues. Les données ainsi collectées sont exploitées pour contrôler la consommation d'énergie des usines connectées.

Si aucune perturbation n'est détectée, l'agent AOE envoie un message indiquant l'absence de la fluctuation de l'énergie aux agents AOU. Si une variation de l'énergie est détectée, l'AOE calcule tout d'abord le taux d'énergie *tauxEnergy %* à réduire. Un ordre *Reschedule* qui précise le temps de ré-ordonnancement et la contrainte énergétique à respecter est alors envoyé. La réception de ce type de message implique que les AOU doivent exécuter une technique de ré-ordonnancement permettant d'obtenir un nouvel ordonnancement adapté aux changements imposés par les fournisseurs d'énergie.

### 3.2.2 Agent Ordonnanceur d'usine (AOU)

L'Agent Ordonnanceur d'usine a pour rôle de :

- Calculer un ordonnancement prédictif moyennant une méthode d'optimisation. (hors ligne)
- Exécuter une technique de ré-ordonnancement suite aux événements aléatoires perçus moyennant des stratégies de négociation. (en ligne)
- Contrôler et détecter localement les perturbations de production. (en ligne)

Dans ce travail, l'ordonnancement prédictif est déterminé par un algorithme de type PSO repris de (**Nouiri et al., 2018)** où une particule représente une solution potentielle au problème. Chacune de ces particules est dotée par un vecteur de position, une vitesse qui permet à la particule de se déplacer et un voisinage c'est-à-dire un ensemble de particules (Neighbours) qui interagissent directement sur la particule, en particulier celle qui a le meilleur critère. La fonction objective eq. (1) utilisée pour évaluer les solutions combine à la fois le makespan ($C_{max}$) et l'énergie totale ($E_{tot}$). Comme les deux objectives ne sont pas de même grandeur ni unité, une phase de normalisation est exigée. Un facteur de pondération γ permet de fixer le degré d'importance de chaque objectif.

$$F = \gamma \frac{C_{max}}{MaxMakespan} + (1 - \gamma) * \frac{E_{tot}}{MaxEnergy} \quad (1)$$

Si, durant la phase prédictive, l'AOU reçoit un message de l'AOE indiquant le refus de sa demande d'énergie issue du calcul initial, l'algorithme PSO est exécuté à nouveau mais avec une nouvelle valeur γ. Ce facteur de pondération sera réduit par une valeur α favorisant ainsi la réduction de la consommation énergétique au détriment de l'efficacité de l'ordonnancement.

La figure suivante illustre le pseudo code du comportement en ligne de l'agent AOU. Il s'agit d'un comportement cyclique lancé périodiquement, la période étant notée p3. Si le message reçu de la part de l'AOE ne fait référence à aucune situation perturbante, l'ordonnancement est appliqué selon le plan initial. Si ce n'est pas le cas, une technique de ré-ordonnancement est exécutée.

```
Control Perturbation (periode p3)
Recevoir ControlMessage de l'agent ordonnanceur d'énergie AOE
- Si (ControlMessage = « no energy perturbation ») alors
    Continuer l'ordonnancement selon la solution prédictive
- Si non
    Extraire la solution perturbée à partir de temps Reschedule.timeResch
    Appliquer la méthode de ré-ordonnancement sur cette sous particule pour trouver un nouveau
ordonnancement qui respecte la contrainte énérgitique Reschedule.tauxE.
```

**Figure 4. Pseudo code du comportement en ligne de l'agent AOU**

Nous détaillons dans ce qui suit les méthodes de ré-ordonnancement utilisées par les AOU en cas de réception d'une alerte de la part de l'AOE indiquant une fluctuation de l'énergie disponible. Plusieurs méthodes de ré-ordonnancement ont été proposées. Par exemple, pour le problème de type job shop classique, nous avons identifié la méthode de décalage à droite, le ré-ordonnancement total et le ré-ordonnancement partiel (**Nouiri et al., 2017**). Avant de présenter la technique de ré-ordonnancement proposée, nous détaillons le type de perturbation considéré.

### 3.3 Type de perturbation considéré

La **figure 5** illustre les différents paramètres composant une perturbation. Générer une perturbation revient à déterminer la ressource affectée, la date d'occurrence de la perturbation, la durée et son type (par exemple, réparable ou non réparable). La ressource affectée peut être soit renouvelable soit consommable. Dans notre travail, nous nous intéressons à la variation de la disponibilité de l'énergie renouvelable comme source d'évènement perturbant dans le système.

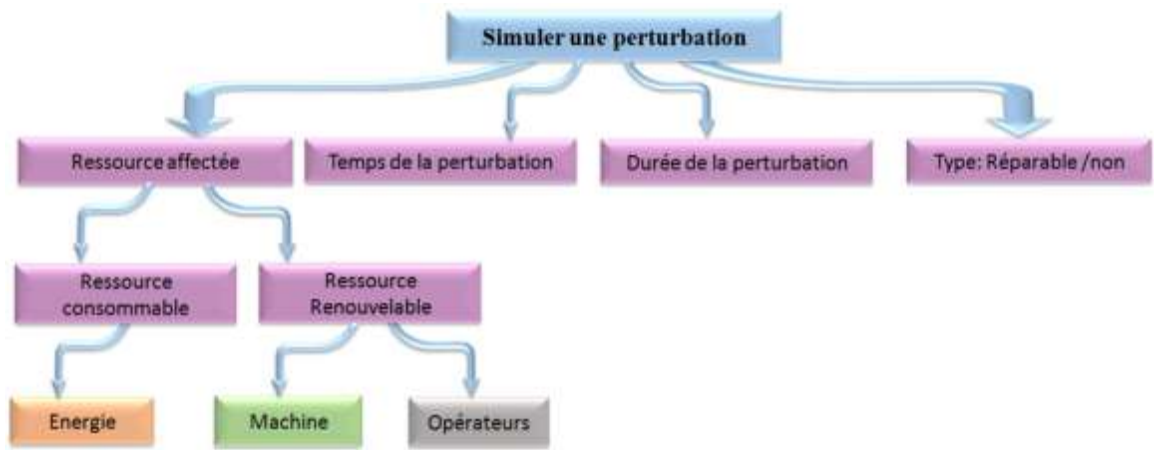

**Figure 5. Types de perturbation**

## 3.4 Les techniques de ré-ordonnancement proposées

Plusieurs techniques de ré-ordonnancement ont été proposées dans la littérature pour résoudre le problème de job shop dynamique. **Vieira et al., (2003)** ont classé ces méthodes en trois groupes : méthode de décalage à droite, ré-ordonnancement total et ré-ordonnancement partiel. Ces méthodes tiennent rarement compte de l'efficience lors de la sélection du nouvel ordonnancement. Dans ce travail, nous avons proposé deux techniques de ré-ordonnancement ayant comme objectif d'assurer l'efficience en respectant un seuil énergétique communiqué en temps réel par les agents AOE. Les techniques de ré-ordonnancement sont exécutées par les agents AOU. Nous rappelons que nous supposons dans cet article que les systèmes de production de ces usines sont de type job-shop. La première étape de la technique de ré-ordonnancement proposée consiste à déterminer les opérations affectées par la panne à partir de l'ordonnancement prédictif. La figure 6 illustre l'organigramme de la première méthode de ré-ordonnancement proposée.

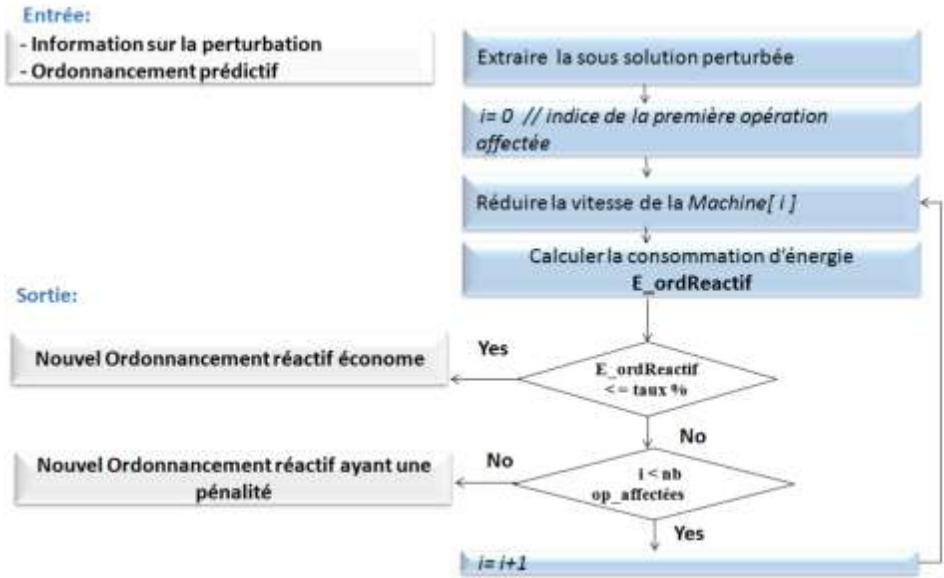

**Figure 6. La technique 1 de ré-ordonnancement proposée**

La méthode est basée sur le changement de vitesse des machines des opérations affectées qui est, nous le rappelons, la variable d'ajustement choisie dans cet article. L'ordre de passage des opérations sur les machines reste identique à celui qui est fixé par l'ordonnancement prédictif afin de maintenir la solution la plus stable possible. À chaque itération, la vitesse de chaque opération affectée est changée et la nouvelle consommation énergétique est alors calculée. Si la nouvelle consommation respecte le taux envoyé par les AOE alors la solution est considérée comme la meilleure. Sinon, ce processus est répété sur toutes les autres opérations affectées. Si à la dernière itération, le taux énergétique demandé n'est pas respecté, alors une pénalité est associée à l'ordonnancement réactif trouvé. Dans ce cas la deuxième méthode de ré-ordonnancement est utilisée. Cette dernière se base sur le changement de séquencements, jusque-là invariants. La solution obtenue risque dès lors d'être très différente de la solution initiale, le résultat risque ainsi d'être moins stable en terme de changements dans la séquence. La figure 7 illustre l'organigramme de la deuxième méthode. L'ordre de passage des opérations aux machines est ainsi ici changé. Dans cette méthode, la recherche s'effectue à partir de permutations entre opérations et les vitesses des machines. À chaque fois, la consommation d'énergie est calculée pour choisir la meilleure solution respectant au mieux la contrainte énergétique.

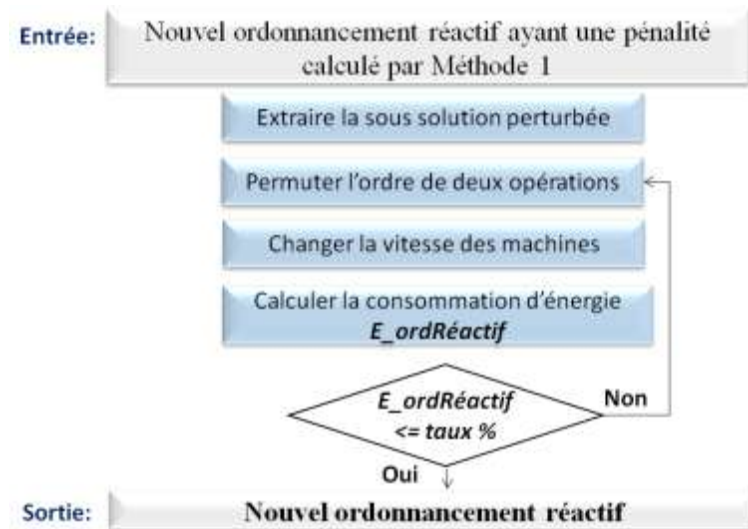

**Figure 7. La technique 2 de ré-ordonnancement proposée**

**4. Implémentation physiquement distribuée de l'architecture multi-agent**

Comme expliqué, l'architecture EasySched est composée de plusieurs AOU, plusieurs AOE et un ACE. La validation par simulation sur un seul calculateur, de manière centralisée est possible. Toutefois, ce type de validation reste partiel. Réaliser une validation réellement distribuée, sur différents systèmes, intégrant les éléments physiques (hardware) et logiciels et capteurs connectés (IOT) est proposé dans cet article. Ces éléments sont connectés en réseaux, ce qui permet un retour d'expérience plus pertinent, notamment dans l'objectif d'appliquer l'architecture sur un système réel, tel que la cellule AIP-PRIMECA de l'Université Polytechnique Hauts-de-France. La figure 8 schématise cette implémentation. L'évolution d'une implémentation centralisée vers l'implémentation proposée dans cet article y est mise en évidence.

Dans un contexte multi agent, la vie d'un agent exige la présence d'un conteneur. Un conteneur est une classe abstraite qui contient tous les services nécessaires à l'hébergement et la gestion des agents durant l'exécution de leur programme. Ces conteneurs d'agents peuvent être répartis sur le réseau. Les agents créés sont par défaut localisés dans le même "Main container". Avant d'être déployé sur le système physiquement distribué composé de systèmes embarqués connectés entre eux, un conteneur est créé pour chaque agent distant. « Usine Container i » et «

Energy Provider Container i » sont ainsi des conteneurs alloués respectivement aux agents AOU, AOE, ACE (**voir la figure 8**). Des systèmes embarqués ARM A7 sont utilisés pour embarquer les containers AOE1, AOE2, AOU1, AOU2 et ACE.

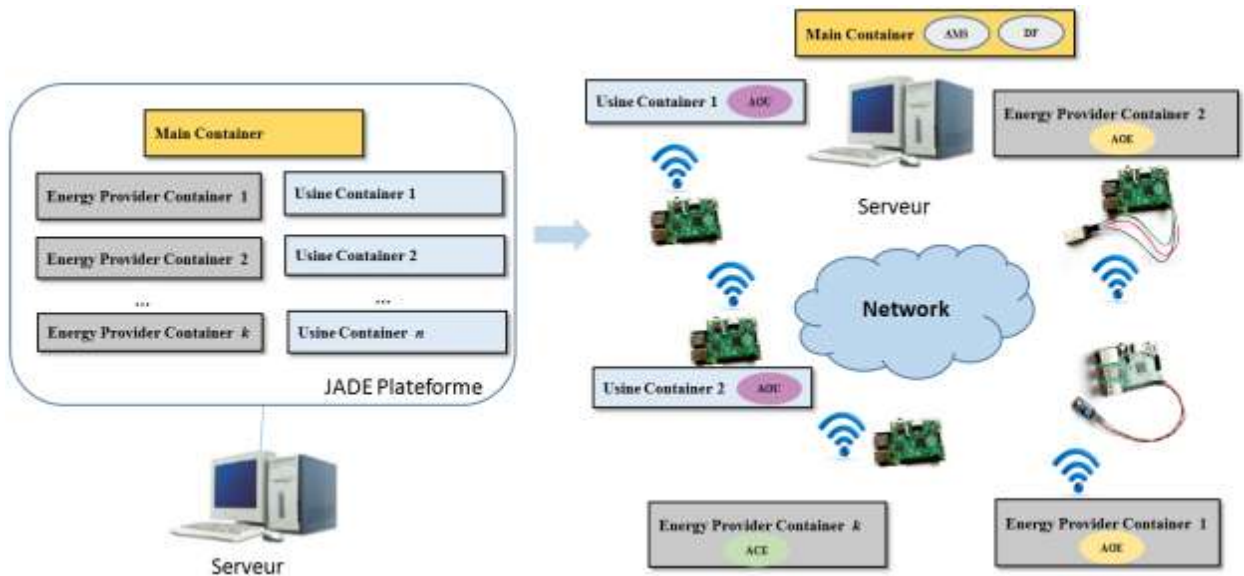

**Figure 8. Implémentation physiquement distribuée de EasySched**

La quantité d'énergie renouvelable reste totalement simulée. Cependant, pour prendre en compte sa forte imprévisibilité, nous avons choisi d'utiliser des capteurs ambiants de température, d'humidité et de luminosité. Les sorties de ces capteurs (qu'il est facile de modifier) représentent le suivi de la variation des grandeurs dont dépend la puissance délivrée par les deux ressources de production d'énergie. Par exemple, la variation de l'éclairement qui influe systématiquement sur la variation de la puissance délivrée par toute centrale photovoltaïque est facilement simulée en masquant plus ou moins le capteur de lumière. De même la puissance délivrée par une centrale éolienne est dépendante de la température et de l'humidité. Elle est modifiable facilement en influençant sur le capteur de température et d'humidité.

La figure suivante contient le pseudo code qui permet aux AOE de détecter une perturbation sur la puissance fournie. Il s'agit d'un comportement cyclique qui se répète chaque période fixée p1.

```
Acquisition donnée (periode p1)
 - lire Temperature & humidité de capteur
 - calcul puissance P(t)
 - Si P(t)/ P(0) < 1 alors
       Alarm="perturbation"
 -Sinon
       Alarm="non perturbation"
```

**Figure 9. Pseudo code du comportement « Acquisition données » d'AOE**

P(t) et P(0) sont les puissances calculées à partir de valeur de température et d'humidité collectés. Dans le contexte de cet article, nous avons choisi d'utiliser le modèle proposé par (**Marshall and Plumb., 2008**), où la puissance électrique délivrée par une éolienne peut être calculée comme suit :

$$P = \frac{1}{2}\rho S V^3$$

Avec S est la surface de l'éolienne, V vitesse de vent et ρ est la masse volumique de l'air calculée comme suit :

$$\rho = \frac{1}{287.06(T+273.15)}\left(p - 230.617 * \varphi * \exp\left(\frac{17.5043 * T}{241.2 + T}\right)\right)$$

avec T et φ sont la température et l'humidité récupéré à partir des capteurs et p = 1 013,25 hPa, la pression de l'air.

Si le rapport P(t) / P(0) est inferieur à 1, la production de l'énergie diminue. La variable "Alarm" déterminant l'état de système est alors mise à jour.

Afin de limiter le bruit de capteur sur les signaux et ainsi éviter de considérer chaque évolution de la valeur des signaux comme une perturbation, nous avons ajouté un algorithme artificiel de filtrage. Pour cela, nous avons contraint le choix de la période de comportement cyclique (p2) plus grande que celle de l'acquisition des données des divers capteurs. La figure suivante contient le pseudo code de ce comportement. La période p2 est alors modifiée quand l'AOE détecte une anomalie (voir figure 10). Ceci a pour premier objectif d'éviter le lancement successif de signaux de ré-ordonnancement et pour second objectif d'espacer les horizons de ré-ordonnancement. La figure suivante 11 illustre la différence entre les différentes périodes.

```
VérificationFausseAlarm(periode p2)
- Si Alarm="perturbation"  alors
      timeResch= timeResch+p2
      Reschedule.timeR=timeResch
      Reschedule.tauxEnergy= (P(0)-P(t)/P(0) )*100
      Envoyer messageObject ( Reschedule, AOU)
      P2=3*P2
-Sinon
      Envoyer message( « no energy perturbation », AOU)
```

**Figure 10. Pseudo code du comportement « vérification Fausse Alarme » d'AOE**

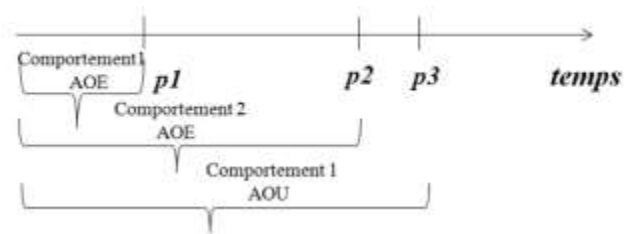

**Figure 11. Valeurs initiales des périodes des comportements cycliques**

## 5. Expérimentations et interprétation

Pour tester l'efficacité et la performance de l'architecture au travers de son implémentation distribuée, nous avons exécuté différents jeux de test construits à partir de trois différents benchmarks de type job-shop composé de machines avec des vitesses variables avec des tailles d'instance (machines x Vmax x Pi) (3x3x10), (3x25x100) et (7x10x100). Les détails des instances sont fournis dans (**Salido et al., 2016**). Chaque instance est exécutée 5 fois. L'implémentation distribuée est constituée d'un PC et de cinq systèmes embarqués ARM A7 (AOE1, AOE2, AOU1, AOU2 et ACE). Le PC dispose d'un processeur Intel "Core2Duo" cadencé à 2, 4 GHz et 8 Go de RAM. La carte intégrée fonctionne à 900 Mhz avec 1Go de RAM. Le système embarqué cible exécute la version Debian de Linux. Toutes les machines exécutent Java SE Embedded. Un LAN est utilisé pour la communication entre tous les composants de l'implémentation distribuée. Pour nos expérimentations, les périodes p1, p2, p3 sont initialement fixées à 3, 6 et 8 s respectivement. La figure 12 contient différentes photos prises de notre implémentation distribuée (au sein du LAMIH).

L'expérimentation est composée de deux étapes : la première permet de valider la partie prédictive de EasySched tandis que la deuxième permet de valider la partie réactive.

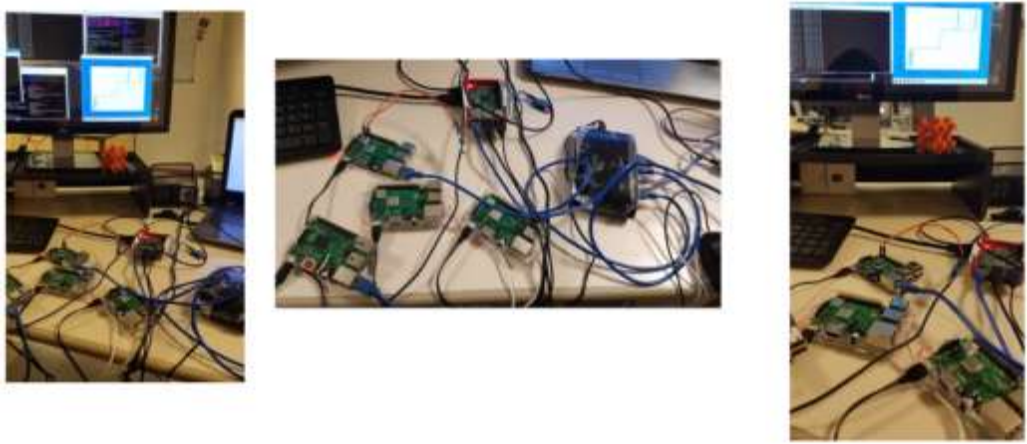

**Figure 12. Implémentation distribuée de EasySched**

## 5. 1 Evaluation de la partie prédictive de l'architecture EasySched

Dans cette partie, nous calculons et évaluons la solution fournie dans la partie prédictive. Le **tableau 1** contient le résultat des essais. La première colonne représentant l'instance à tester, la deuxième la qualité de la solution fournie par l'AOU1, AOU2 en termes de Makespan (Mk, en s) et l'énergie consommée (E en wh), tout en précisent la valeur du facteur de pondération γ et le message envoyé par le AOE1 et 2 (deux états : oui ou non) qui permettra d'arrêter ou non la recherche prédictive de la solution. Chaque AOU envoie une demande de besoin énergétique après avoir calculé un ordonnancement prédictif en utilisant le PSO. Si l'AOE dispose de suffisamment d'énergie, celui-ci va envoyer un message « Oui » indiquant qu'il est possible de procéder avec cet ordonnancement. Si non, il envoie un message « Non » à l'agent AOU pour forcer le calcul d'une autre solution. Dans ce cas, l'AOU va exécuter le PSO de nouveau mais en réduisant la valeur du facteur de pondération, ce qui favorisera une réduction de la consommation d'énergie jusqu'à l'obtention d'une solution satisfaisante, correspondante à l'envoi du message « Oui ».

Ces résultats illustrent le bon fonctionnement du mécanisme de coopération qui se met en place entre les consommateurs et les fournisseurs d'énergie pour converger vers des ordonnancements acceptables. Bien évidemment, d'autres tests sont à mener pour conclure de manière plus crédible sur ce bon fonctionnement. Des comparaisons avec les méthodes existantes doivent également être menées.

**Tableau 1 : Résultats- partie prédictive**

| Instance | AOU 1 | | | AOE1 | AOU 2 | | | AOE1 |
|---|---|---|---|---|---|---|---|---|
| | γ | Mk | E | Message envoyé | γ | Mk | E | Message envoyé |
| 3x5x10 | 1 | 41 | 145.8 | Non | 1 | 41 | 145.8 | Non |
| | 0.9 | 41 | 133.7 | Non | 0.9 | 41 | 133.2 | Oui |
| | 0.8 | 42.2 | 123.5 | Non | - | - | - | - |
| | 0.7 | 45.3 | 112.6 | Oui | - | - | - | - |
| 7x10x100 | 1 | 625.9 | 2773.4 | Non | 1 | 625.9 | 2773.4 | Non |
| | 0.9 | 626 | 2560.7 | Oui | 0.9 | 626 | 2560.7 | Non |
| | - | - | - | - | 0.8 | 642.4 | 2418.9 | Oui |
| 3x25x100 | 1 | 1711.8 | 6797.2 | Non | 1 | 1711.8 | 6797.2 | Non |
| | 0.9 | 1732.2 | 6311 | Non | 0.9 | 1732.2 | 6311 | Non |
| | 0.8 | 1791.4 | 5726.2 | Non | 0.8 | 1791.4 | 5726.2 | Non |
| | 0.7 | 1943.7 | 5097.4 | Non | 0.7 | 1886.3 | 6107.7 | Non |

| | 0.6 | 2118.3 | 4617.4 | Non | 0.6 | 2197.6 | 4547.9 | Non |
| | 0.5 | - | - | - | 0.5 | 2317.1 | 4257.3 | Oui |

## 5. 2 Evaluation de la partie réactive de l'architecture EasySched

Dans cette partie, nous testons le comportement réactif de l'architecture EasySched face à une fluctuation de l'énergie renouvelable disponible chez les AOE. La figure suivante illustre la solution prédictive délivrée par l'AOU1 connecté avec le fournisseur d'énergie que nous utiliserons. Le makespan trouvé est égal à 44 s avec une consommation d'énergie égale à 133 Wh. La particule (PSO) correspondante à cette solution ainsi que le diagramme de Gantt sont donnés dans la figure 13.

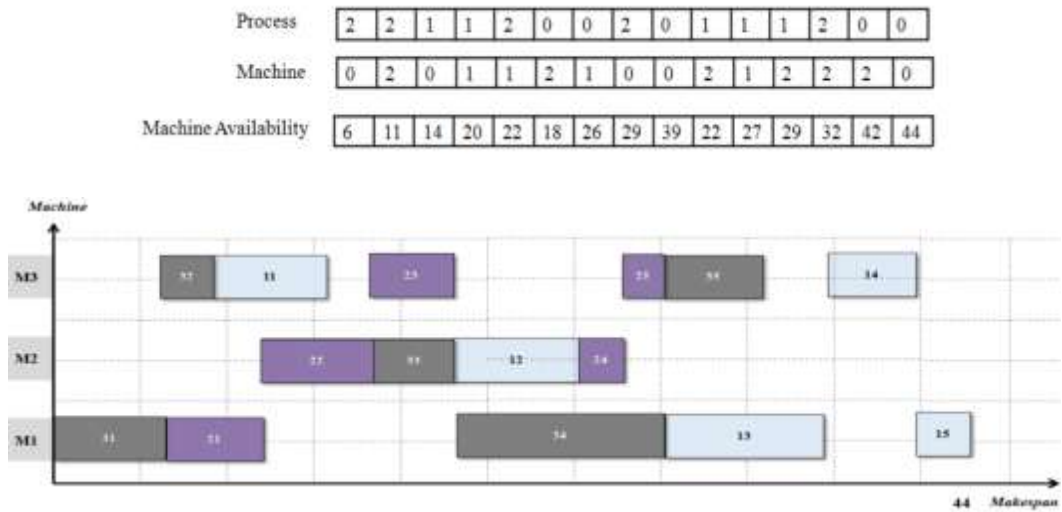

**Figure 13. Diagramme de Gantt de l'ordonnancement prédictif trouvé par AOU1**

La figure 14 contient la courbe de la variation de température et d'humidité collectée par l'AOE à partir des capteurs nous permettant de simuler en temps réel et de manière complètement imprévisible les conditions météorologiques. Par exemple, à l'instant 22 s, la valeur de la température est « 28 » et celle de l'humidité « 33 ». La valeur de la puissance à cet instant est alors plus petite que celle à l'instant initial (T=19, H=36). Le rapport de P(t)/P(0) est égal à 0,711. Par conséquent, une alerte est délivrée aux AOU indiquant qu'il faut recalculer une nouvelle solution présentant une consommation d'énergie réduite de 26%.

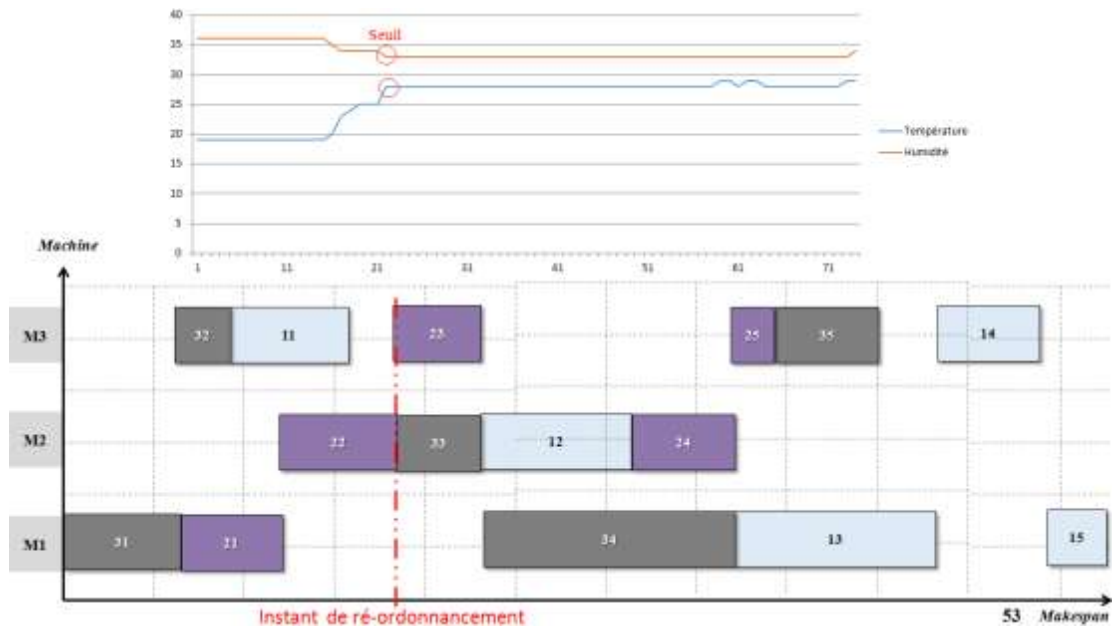

**Figure 14. Diagramme de Gantt de l'ordonnancement réactif trouvé par AOU1**

La consommation d'énergie de l'ordonnancement réactif trouvé est désormais réduite pour être égale à 94Wh après la perturbation alors que le makespan se dégrade et vaut désormais 53 s. Dans ce scenario, l'adaptation de la consommation d'énergie des usines pour respecter le seuil énergétique 26% a engendré une dégradation de performance sous la forme d'une augmentation de la date de réalisation de la production. Il faut noter que même si l'on peut s'attendre à ce que ce type de comportement se reproduise dans d'autres situations (dégradation du makespan pour tenir compte d'une plus faible disponibilité énergétique), la quantification de cette dégradation dépend des modèles utilisés pour la production et la consommation d'énergie.

Cette solution est trouvée en utilisant la première méthode de ré-ordonnancement proposée qui cherche à préserver la stabilité et ne changer que les vitesses des machines jusqu'à satisfaire la contrainte énergétique. Les vitesses des machines M1, M2, M3 sont ainsi changées à partir de l'instant de ré-ordonnancement. Le tableau suivant contient d'autres résultats de simulation. Les taux d'énergie délivrés par les AOE sont précisés. Dans notre cas, le pourcentage envoyé par l'AOE photovoltaïque est fixe et égal à 10 %. L'expérimentation se conclut ainsi favorablement, la réactivité de notre architecture multi-agent étant mise en évidence sur ce cas d'étude.

Tout comme pour la partie prédictive, des études complémentaires doivent être menées au niveau réactif pour pouvoir conclure de manière plus générale sur l'efficacité de la méthode proposée. Des protocoles expérimentaux rigoureux sont en cours de développement. Des tests statistiques devront être menés. Cependant, les résultats obtenus et présentés dans cet article nous confortent dans cette efficacité.

**Tableau 2 : Résultats- partie réactive**

| Instance | AOU 1 | | | AOE1 | AOU 2 | | | AOE1 |
|---|---|---|---|---|---|---|---|---|
| | Initial Mk, E | Mk | E | Taux E% | Initial Mk, E | Mk | E | Taux E% |
| 3x5x10 | 44, 133 | 53 | 94 | 26 | 41,145.8 | 42 | 123 | 10 |
| 7x10x100 | 626,2773 | 703 | 2099 | 22 | 626,2773 | 642 | 2418 | 10 |
| 3x25x100 | 1711, 6797 | 2056 | 4742 | 28 | 1711,6797 | 1854 | 5466 | 10 |

## 6. Conclusion

Dans ce travail nous avons proposé EasySched, une architecture multi-agent pour l'ordonnancement prédictif et réactif de la production en fonction d'une énergie renouvelable disponible difficilement prévisible. Nous avons proposé une implémentation distribuée originale de l'architecture qui permet de valider ces mécanismes sur des instances issues de la littérature. Une première perspective concerne l'intégration d'un mécanisme permettant de basculer entre les différentes sources d'énergie disponible au travers d'un pourcentage déterminé par négociation entre agents. Une deuxième perspective consiste à proposer une architecture holonique multicouche pour généraliser EasySched. Dans ce cas, l'agent ordonnanceur représentant l'usine connecté serait un holon récursivement composé par un ensemble d'holons représentant chacun une machine ou un produit intelligent dans l'industrie.